\documentclass[reprint,aps,groupedaddress,showpacs]{revtex4-1}
\usepackage{graphicx}
\usepackage{amsmath,amssymb,lmodern}
\usepackage{amsfonts}
\usepackage{amssymb}
\usepackage{bm}

\begin{document}

\title{Topological hybrid semimetal phases and anomalous Hall effects in a three dimensional magnetic topological insulator}

\author{M. N. Chen$^1$}
\email{mnchen@hdu.edu.cn}

\author{W. C. Chen$^{2}$}

\author{Yu Zhou$^{1}$}

\affiliation{$^1$School of Science, Hangzhou Dianzi University, Hangzhou, 310018, China
\\
$^2$Institute of Advanced Magnetic Materials, College of Materials and Environmental Engineering, Hangzhou Dianzi University, Hangzhou, 310018, China
}

\begin{abstract}
  In this work, we propose a ferromagnetic Bi$_2$Se$_3$ as a candidate to hold the coexistence of
  Weyl- and nodal-line semimetal phases, which breaks the time reversal symmetry. We demonstrate that the type-I Weyl semimetal phase, type-I-, type-II- and their hybrid nodal-line semimetal phases can arise by tuning the Zeeman exchange field strength and the Fermi velocity. Their topological responses under U(1) gauge field are also discussed. Our results raise a new way for realizing Weyl and nodal-line semimetals and will be helpful in understanding the topological transport phenomena in three-dimensional material systems.
\end{abstract}

\maketitle



\section{\label{section1}INTRODUCTION}
In recent decades, three-dimensional (3D) 
topological semimetals (TSs)~\cite{Zhang1} and their exotic transport phenomena
have attracted a great surge of research interests. 3D TSs, including Dirac-~\cite{Young2},
Weyl-~\cite{Wan3,Weyl2,Weyl3,Weyl4,Weyl5,Weyl6} and nodal-line semimetals
(NLSMs)~\cite{Morimoto4,Zhao5,Zhao6,Sigrist7}, exhibit a variety of interesting features
in their electronic and transport properties,
such as surface Fermi arc states, chiral magnetotransport and negative magnetoresistance~\cite{Dai8,3DTI1}.
Unlike topological insulators, the TSs host topologically nontrivial
bulk electronic structures with several band-contact points
or lines, which are protected by certain discrete symmetry~\cite{Zhao10}.
According to the type of band dispersion around the touching points,
the Weyl semimetals (WSMs) can be classified into type-I and type-II, in which
their transport properties are qualitatively different. While they still have strong topological responses ascribed to the gapless band structures, the corresponding topological invariants are defined in terms of the Berry bundle that encloses the
Fermi surface from its transverse dimension,
rather than in the whole Brillouin zone.

Following the Dirac semimetals and the WSMs,
the NLSMs represent a new type of TSs,
whose bands can cross each other along a closed curve.
Similarly, just like the WSMs,
the NLSMs can also be classified into type-I and type-II categories.
For type-II NLSMs~\cite{He11},
the zero energy bulk states have a closed loop in momentum space
but the (local) Dirac cones on the node line become tilted~\cite{Zhao12,Zhang13,Hyart14,Pisani15}.
In the past few years, much effort has been devoted to searching
for NLSMs theoretically in many contexts, such as the solid-state, atomic,
photonic and acoustic systems~\cite{offresonant,cold1,photonic1,
offresonant, Silicene,nodal1,nodal2,nodal3,nodal4,nodal5,nodal6,
nodal7,nodal8,nodal9,Bomantara2016, Wang2014,Zhang2016, Zhou2016, Wang2018, Bomantara2018, Na3Bi}.
Recently, first-principle calculations have revealed Be$_2$Si is a candidate of type-I and type-II hybrid NLSMs~\cite{Li11} in non-magnetic system.
A toy model based on the modified Dirac equation is also proposed
for the ferromagnetic WSMs and NLSMs~\cite{Rauch2017}.
However, experimental realization and observation of NLSMs 
in real 3D ferromagnetic materials remain a challenge~\cite{Ekahana1,Neupane2,Li3,Bian4,Wu5,Hu6}.
So far, the theoretical proposals of ferromagnetic NLSMs
have still remained elusive because Zeeman exchange field caused
by ferromagnetism could open a gap ascribed to time-reversal symmetry (TRS) breaking. 
It seems that from the point of symmetry protected mechanisms
for NLSMs the Weyl semimetal phase can emerge in a ferromagnetic system while NLSMs can not.
Therefore, a natural question on whether WSM and NLSM can coexist in a ferromagnetic system remains to be worthy of study.

To this end, it is highly desired to study whether NLSMs can emerge in a ferromagnetic system.
In this work, we show that a ferromagnetic Bi$_2$Se$_3$ is a candidate to hold the coexistence of
WSM and ferromagnetic NLSM phases. With the increasing
strength of Zeeman exchange field $g_\mathrm{z}$,
it is found that such system features the WSM phases, type-I-, type-II NLSM
as well as type-I and type-II hybrid NLSM phases.
We also discussed their topological
responses under U(1) gauge field. 

This paper is organized as follows. In Sec.\ref{section2},
we analytically solve the model Hamiltonian and demonstrate the existence of WSM phase
and type-I, type-II and their hybrid NLSM phases for distinct parameters. In Sec.\ref{symmetry},
we analyze the symmetries and topological invariants for our system. In Sec.\ref{section3},
we calculate the anomalous Hall conductivity
and discuss the effects of WSMs and NLSMs on these transport
quantities. In Sec.\ref{field}, we discuss effective action for electron states
which are closed to the nodal-line.
The last section contains a conclusion.

\section{\label{section2}The Model}

Let us begin by considering the low-energy $\mathbf{k}\cdot\mathbf{p}$
effective Hamiltonian
\begin{equation}
H_{0}=\epsilon_{\mathbf{k}}\hat{\sigma}_0+A_1k_z\hat{\tau}_x\hat{\sigma%
}_z+A_2\hat{\tau}_x(k_x\hat{\sigma}_x+k_y\hat{\sigma}_y) +M_{
\mathbf{k}}\hat{\tau}_z\ ,
\label{ham1}
\end{equation}
which was used to describe the bulk state of Bi$_2$Se$_3$~\cite{Zhang1}. Here,
$\hat{\sigma}_i$ and $\hat{\tau}_i$ ($i=0$, $x$, $y$, or $z$) are the
Pauli matrices acting in spin and orbital spaces, respectively, $\epsilon_{\mathbf{k}}=C+D_1k^2_z+D_2\mathbf{k}
^2_\parallel$ is the kinetic part, and $M_{\mathbf{k}
}=M_0-B_1k_z^2-B_2\mathbf{k}_\parallel^2$ with $\mathbf{k}
_\parallel^2=k_x^2+k_y^2$ is the Dirac mass term expanded to the second
order of the momentum. Particularly, this model describes a 3D strong TI when $M_0B_1>0$ and $M_0B_2>0$.
A Zeeman exchange field can be created by doping of magnetic atoms,
such as Mn, or depositing a ferromagnetic insulator on top or bottom surface of 3D TI,
which breaks time-reversal symmetry. The Zeeman exchange field is along the $z$-axis and can be
described by $H_{\mathrm{exc}}=g_{\mathrm{z}}\hat{\sigma}_z$,
where $g_{\mathrm{z}}$ is the Zeeman coupling strength.
The full Hamiltonian then reads $H=H_0+H_{\mathrm{exc}}$.
Making a unitary transformation $\mathcal{H}=U^\dag HU$ with
the unitary operator
\begin{equation}
U=\frac{1}{2}(1+\hat{\tau}_x)+\frac{1}{2}(1-\hat{\tau}_x)\hat{\sigma}_z\ ,
\end{equation}
this yields
\begin{equation}
\mathcal{H}(\mathbf{k})=\epsilon_{\mathbf{k}}\hat{\sigma}_0+A_2(k_x\hat{\sigma}
_x+k_y\hat{\sigma}_y) +h_\tau(\mathbf{k})\hat{\sigma}_z\ ,
\label{ham2}
\end{equation}
where $h_\tau(\mathbf{k})=M_{\mathbf{k}}\hat{\tau}_z+A_1k_z\hat{\tau}_x+g_{\mathrm{z}}\hat{\tau}_0$,
which can be regarded as a sub-Hamiltonian in $\hat{\tau}$-space. In fact, we can find that
$h_\tau(\mathbf{k})$ is a minimal Hamiltonian for a NLSM if we neglect the constant term $g_{\mathrm{z}}$.
We can also find that the degrees $\hat{\tau}_i$ and $\hat{\sigma}_i$ have
totally been separated through the unitary transformation.

\begin{figure*}
  \centering
  \includegraphics[width=7.2in]{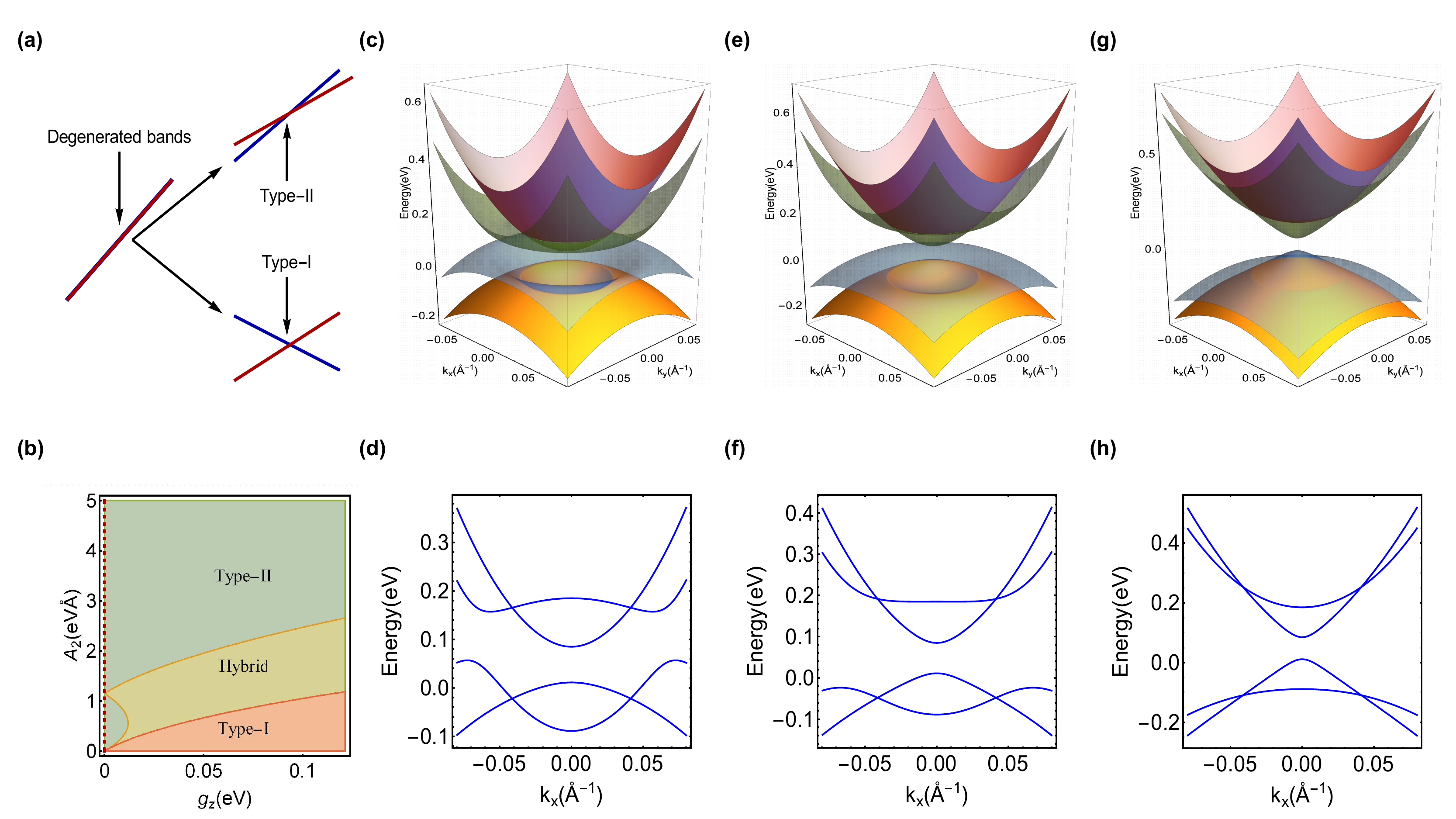}
  \caption{The Phase diagram and corresponding band structures for varied parameters,
  (a) The schematic of band structures for type-I and type-II NLSMs.
  (b) The Phase diagram as functions of Fermi velocity $A_2$(eV$\cdot${\AA}) and Zeeman exchange field $g_\mathrm{z}$(eV).
  (c) Band structure for type-I NLSMs and
  (d) corresponding section graph of the band structure (c) when Fermi velocity $A_2=0.853$eV$\cdot${\AA}.
  (e) Band structure for hybrid type-I and type-II NLSMs and
  (f) corresponding section graph of the band structure (e) when Fermi velocity $A_2=2$eV$\cdot${\AA}.
  (g) Band structure for type-II NLSMs and
  (h) corresponding section graph of the band structure
  (g) when Fermi velocity $A_2=3.853$eV $\cdot${\AA}. Other parameters $M_0=-0.05$eV, $C=0.048$eV, 
  $D_1=1.409$eV$\cdot${\AA}$^2$, $D_2=13.9$eV$\cdot${\AA}$^2$,
  $B_1=-3.351$eV$\cdot${\AA}$^2$, $B_2=-29.36$eV$\cdot${\AA}$^2$, $A_1=2.512$eV$\cdot${\AA}
  and $g_{\mathrm{z}}=0.087$eV are fixed.}\label{Band}
\end{figure*}

The eigenvalues and eigenstates of Hamiltonian (\ref{ham2}) can be obtained through a
straightforward diagonlization, and we then obtain the eigenvalues
\begin{equation}
E^{\mathrm{v}/\mathrm{c}}_{\kappa}(\mathbf{k})=\epsilon_{\mathbf{k}}\mp
\sqrt{A^2_2\mathbf{k}^2_\parallel+\Delta^2_{\kappa}(\mathbf{k})}
\end{equation}
and eigenstates
\begin{equation}
|\psi^{\mu}_\kappa(\mathbf{k})\rangle
=|\phi_\kappa(\mathbf{k})\rangle\otimes|\chi^{\mu}_\kappa(\mathbf{k})\rangle
\end{equation}
with superscript $\mu=\mathrm{v},\mathrm{c}$ representing the valence
and conduction bands, respectively. Here, $|\phi_+(\mathbf{k})\rangle=\big[\mathrm{sgn}(A_1k_z)\cos\theta_{\mathbf{k}
},\sin\theta_{\mathbf{k}}\big]^{\textsf{T}}$ and $|\phi_-(\mathbf{k})\rangle=\big[
\mathrm{sgn}(A_1k_z)\sin\theta_{\mathbf{k}},-\cos\theta_{\mathbf{k}
}\big]^{\textsf{T}}$, which represent the eigenvectors for the operator in the square bracket in equation (\ref{ham1}),
where $2\theta_{\mathbf{k}}=\mathrm{arccot}(M_\mathbf{k}/|A_1k_z|)$.
$|\chi^{\mathrm{v}}_\kappa(\mathbf{k})\rangle=(\mathrm{e}^{-\mathrm{i}\varphi_{\mathbf{k}_\parallel}}\sin\alpha_{\mathbf{k}
\kappa},-\cos\alpha_{\mathbf{k}\kappa})^{\textsf{T}}$ and $|\chi^{\mathrm{c}
}_\kappa(\mathbf{k})\rangle=(\mathrm{e}^{-\mathrm{i}\varphi_{\mathbf{k}_\parallel}}\cos\alpha_{
\mathbf{k}\kappa},\sin\alpha_{\mathbf{k}\kappa})^{\textsf{T}}$ are
eigenvectors for valence and conductance bands, $\mathrm{e}^{\mathrm{i}\varphi_{\mathbf{k}
_\parallel}}=(k_x+\mathrm{i}k_y)/|\mathbf{k}_\parallel|$ and $2\alpha_{\mathbf{k}
\kappa}=\mathrm{arccot}[\Delta_\kappa(\mathbf{k})/(A_2|\mathbf{k}
_\parallel|)]$ with $\Delta_{\kappa}(\mathbf{k})=g_{\mathrm{z}}+\kappa\lambda(
\mathbf{k})$, $\kappa=\pm$ and $\lambda(\mathbf{k})=\sqrt{M_\mathbf{k}^2+A^2_1k^2_z}$.
When the Zeeman splitting strength $g_{\mathrm{z}}$ increases,
the degeneracy of bands is lifted,
except the location of the nodal-ring,
which is shown in Fig.\ref{Band}a.
The line degeneracy is protected by a combined $\mathcal{PT}$-symmetry.
Here, we refer the $\mathcal{T}$-symmetry as a pseudo time-reversal symmetry (TRS),
which is related to the pesudo-spin degree of freedom $\hat{\tau}_i$. The real TRS is broken by
the $H_{\mathrm{exc}}$, obviously.

In fact, we can find the location of nodal-ring
in momentum space by solving the equation $E^\mu_+(\mathbf{k})-E^\mu_-(\mathbf{k})=0$ in the $k_x$-$k_y$ plane (i.e. $k_z=0$),
which yields the radius of the nodal-ring as $k^{\mathrm{NL}}_\parallel=\sqrt{M_0/B_2}$.
We can write the momentum as $\mathbf{k}=\pm\mathbf{k}^{\mathrm{NL}}_\parallel+\mathbf{p}$
and expand the Hamiltonian to linear order of $\mathbf{p}$, yielding the
low energy effective Hamiltonian close to the nodal-line:
\begin{equation}
\mathcal{H}^\chi_{\mathrm{NL}}(\mathbf{p})
=h^\phi_\sigma(\mathbf{p})+h^\chi_\tau(\mathbf{p})\hat{\sigma}_z\ ,
\label{nlsm1}
\end{equation}
where
\begin{equation}
h^\phi_\sigma(\mathbf{p})=A_2p_\parallel\hat{\tau}_0(\cos\phi\hat{\sigma}_x+\sin\phi\hat{\sigma}_y)
+g_\mathrm{z}\hat{\sigma}_z\ ,
\end{equation}
\begin{equation}
h^\chi_\tau(\mathbf{p})=-\chi A_3p_\parallel\hat{\tau}_z+A_1p_z\hat{\tau}_x\ ,
\end{equation}
$\mathrm{e}^{\mathrm{i}\phi}=(p_x+\mathrm{i}p_y)/p_\|$ with $p_\parallel=\sqrt{p^2_x+p^2_y}$ being the polar angle,
$\chi=\pm1$ denotes the situation of the expansion at $\pm\mathbf{k}^{\mathrm{NL}}_\parallel$,
and $A_3=2\sqrt{M_0B_2}$.
We can find that the set of parameters $(p_\|,\phi,p_z)$ forms a cylindrical
coordinates system, thus the Hamiltonian (\ref{ham2}) is decomposed into
a family of (2+1)-dimensional subsystems parameterized by $\phi$.
The types of NLSM can be tuned by varying the values of parameters $A_2$ and $g_z$,
the corresponding phase diagram and band structures are shown in Fig.\ref{Band}b-\ref{Band}h.
In Sec.\ref{field}, we will see that the Hamiltonian (\ref{nlsm1}) indeed corresponds
to the Chern-Simons effective action.

\begin{figure}
  \centering
  \includegraphics[width=3.5in]{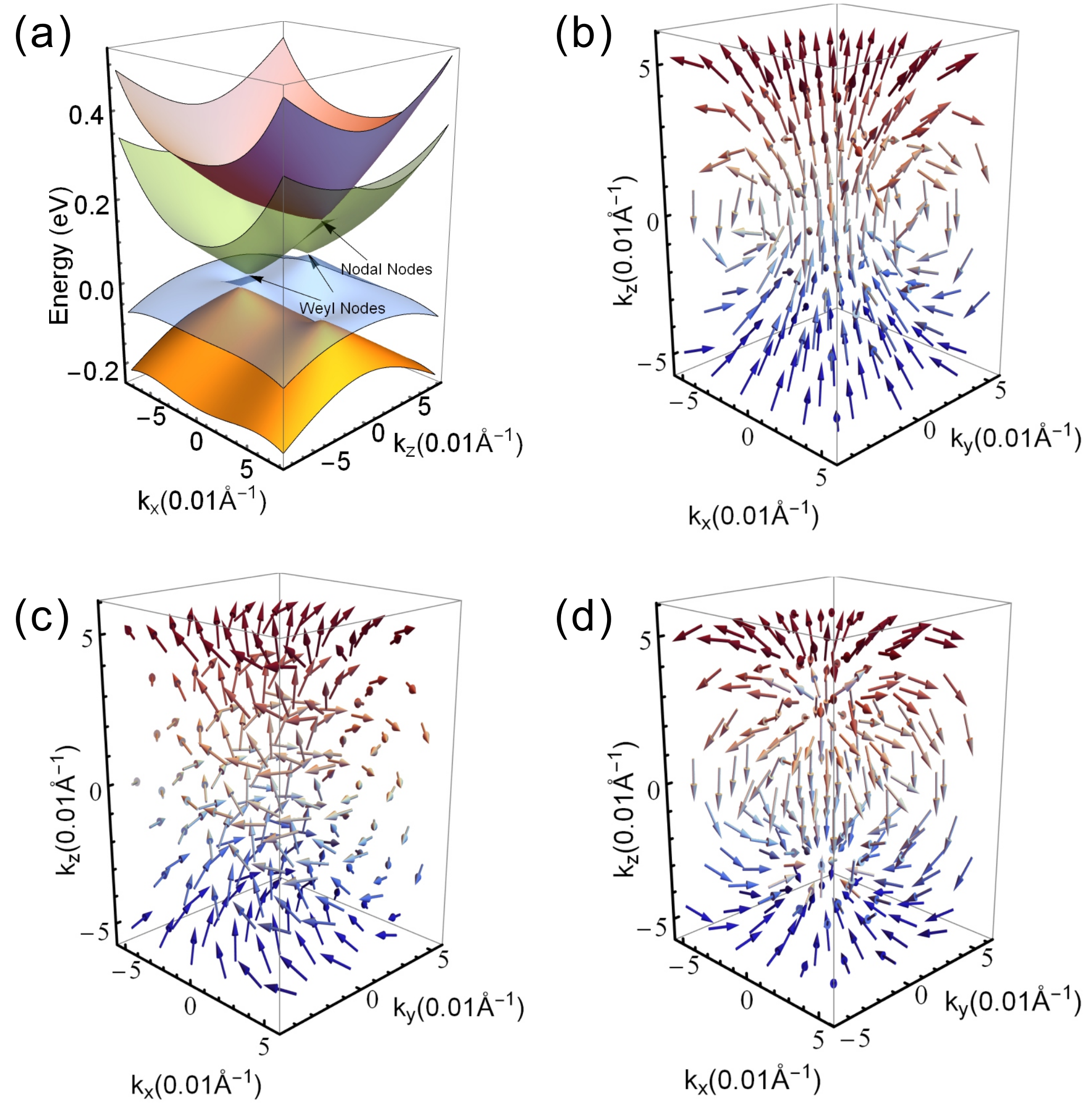}
  \caption{(a) 3D plot of the energy dispersion as the functions of $k_x$ and $k_z$ ($k_y=0$).
  (b)-(d) Vector plots of the Berry curvature for the band with $\mu=\mathrm{v}$ and $\kappa=-$
  for parameters ($\Delta=0$eV, $g_{\mathrm{z}}=0.015$eV), ($\Delta=0.05$eV, $g_{\mathrm{z}}=0.015$eV) and
  ($\Delta=0$eV, $g_{\mathrm{z}}=0.057$eV).}
  \label{Berry}
\end{figure}

Now, let us turn to the discussions on the Weyl semimetal phase.
The solution of equation $E^\mathrm{v}_-(\mathbf{k})-E^\mathrm{c}_-(\mathbf{k})=0$
when $k_x=k_y=0$ gives the locations of Weyl nodes in momentum space as $(0,0,\pm k^\mathrm{w}_z)$, where
\begin{equation}
k^\mathrm{w}_z=\big[(\alpha^2+\beta)^{\frac{1}{2}}-\alpha\big]^{\frac{1}{2}}\ ,
\label{WeylNodes}
\end{equation}
with $\alpha=\frac{A^2_1}{2B^2_1}-\frac{M_0}{B_1}$ and $\beta=\frac{g^2_{\mathrm{z}}-M^2_0}{B^2_1}$.
The equation (\ref{WeylNodes}) indicates that the Weyl nodes can emerge only if $g_{\mathrm{z}}>|M_0|$,
coexisting with the nodal-line, as illustrated in Fig.\ref{Berry}a.
Expanding the Hamiltonian (\ref{ham1}) to linear order of $\mathbf{k}$ around Weyl nodes, one can obtain the standard
Weyl Hamiltonian
\begin{equation}
\mathcal{H}_{\mathrm{Weyl}}^\chi(\mathbf{k})=v_xk_x\hat{\sigma}_x+v_yk_y\hat{\sigma}_y+\chi v_zk_z\hat{\sigma}_z\ ,
\end{equation}
where $\chi=\pm$ denotes the chirality for two Weyl nodes, respectively, $v_x=v_y=A_2$, and
$v_z=\frac{2B^2_1}{g_\mathrm{z}}[(\alpha^2+\beta)^{\frac{3}{2}}-\alpha(\alpha^2+\beta)]^{\frac{1}{2}}$.

For bulk Bi$_2$Se$_3$ material, we usually have $M_0=-0.169$eV,
which is much lager than the typical strength of the exchange field induced
by a ferromagnetic insulator. A recent experimental work shows that suitable atomic
doping can effectively tune the bulk gap of Bi$_2$Se$_3$~\cite{3DTI1},
which enables the Zeeman exchange field to close the bulk gap and host the Weyl nodes.
Here, in our numerical calculations we will use the parameters as $M_0=-0.05$eV and $g_{\mathrm{z}}=0.087$eV,
other parameters are taken the same as Ref.\onlinecite{Dai8}.

\section{Symmetry analysis and topological invariant}\label{symmetry}

\begin{figure}
  \centering
  \includegraphics[width=2.9in]{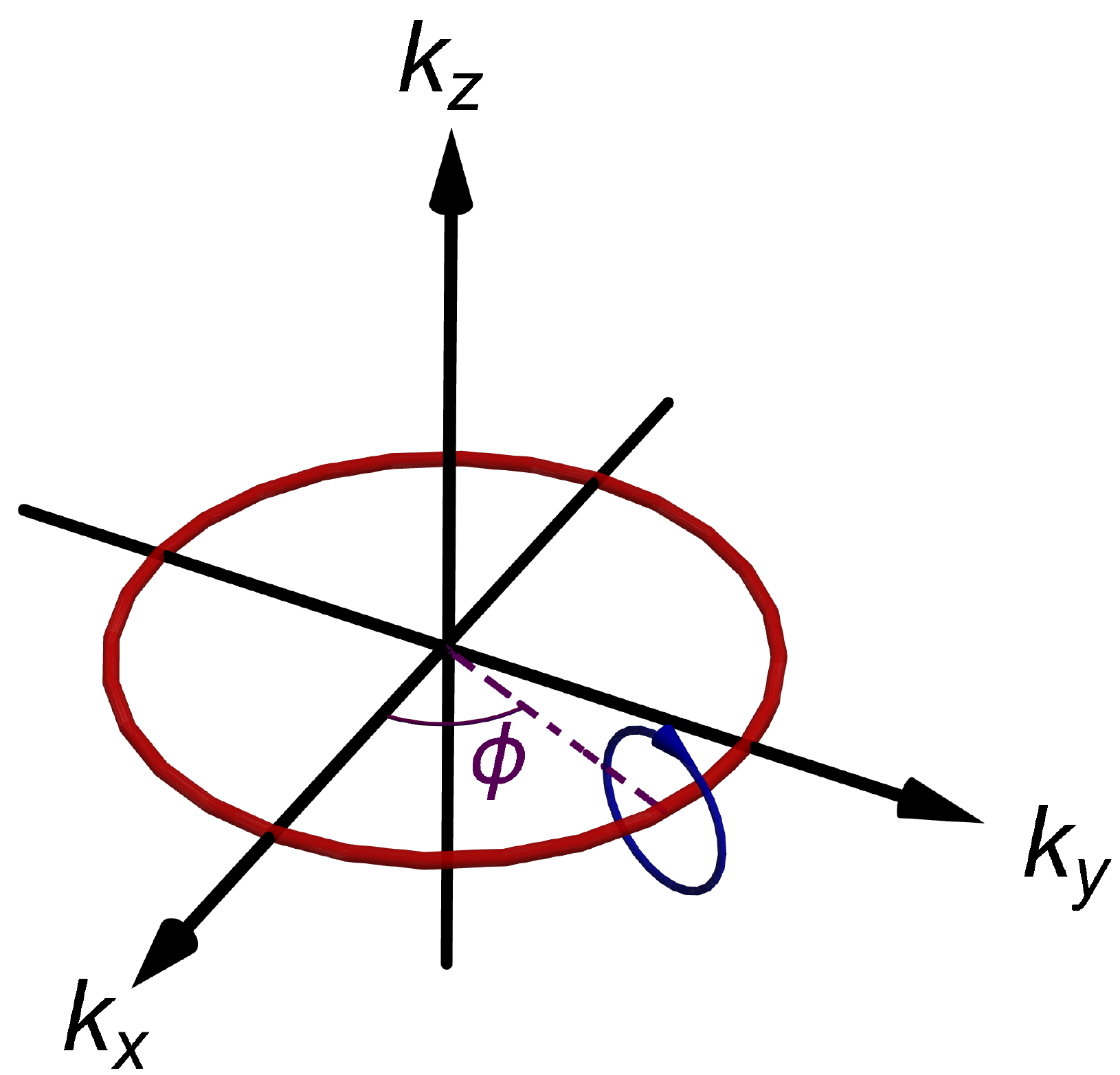}
  \caption{The nodal-line (red) and the integral loop (blue) for the Berry phase.}\label{nodal}

\end{figure}
Symmetries usually play a crucial role in the investigation of topological semimetals.
We can find that the Hamiltonian $\mathcal{H}(\mathbf{k})$ is TRS-breaking while the sub Hamiltonian
$h_\tau(\mathbf{k})$ preserves the pseudo TRS, inversion,
and SU(2) pseudospin-rotation symmetries,
which are corresponding to the pseudospin $\hat{\tau}_i$, and
the combined symmetry of pseudo time-reversal operator $\mathcal{T}=\hat{\tau}_z\hat{\mathcal{K}}$
and parity operator $\mathcal{P}=\hat{\tau}_z$,
i.e. pseudo $\mathcal{PT}$ symmetry with the $\mathcal{PT}$ operator
$\mathcal{\hat{P}\hat{T}}=\hat{\sigma}_0\otimes\hat{\tau}_0\hat{\mathcal{K}}$,
where $\hat{\mathcal{K}}$ is the complex-conjugate operator~\cite{fang}. The $\mathcal{H}$
also has a mirror symmetry with the operator $\hat{\mathcal{M}}_z=\hat{\sigma}_0\otimes\hat{\tau}_z$,
i.e. $\hat{\mathcal{M}}_z\mathcal{H}(k_x,k_y,-k_z)\hat{\mathcal{M}}^{-1}_z=\mathcal{H}(k_x,k_y,k_z)$,
for which the nodal-ring is confined within the plane $k_z=0$.
These two symmetries are in fact independent,
which means that the nodal-ring is robust as long as one of them is preserved.
The $\mathcal{PT}$ symmetry requires the $\pi$ Berry phase, which leads to
a $\mathbb{Z}_2$ topological charge $\nu$, whose form is given by
\begin{equation}
\nu[\mathcal{L}]=\frac{1}{\pi}\int_{\mathcal{L}}\mathrm{d}\varphi\,\mathrm{tr}[A(\varphi)]
~\mathrm{mod}~2\ ,
\end{equation}
where the integration path is along the closed loop $\mathcal{L}$
and the trace only counts the occupied states, as is shown in Fig.\ref{nodal}.
The $\nu$ can be obtained by the Wilson loop, whose form is given as
$\gamma_{\mathcal{L}}=\mathrm{i}\ln W_{\mathcal{L}}$, where
\begin{equation}
W_{\mathcal{L}}=\hat{P}\,\mathrm{e}^{\mathrm{i}\int_{\mathcal{L}}\mathrm{d}\mathbf{l}\cdot\mathbf{A}(\mathbf{k})}
\end{equation}
with the path-ordering operator $\hat{P}$ and Berry connection
$\mathbf{A}^{\mu}_{\kappa\kappa'}(\mathbf{k})=\mathrm{i}\langle\psi^\mu_{\kappa}(\mathbf{k})|
\partial_\mathbf{k}|\psi^\mu_{\kappa'}(\mathbf{k})\rangle$, which is a matrix with dimension equal
to the number of occupied bands.
Loops $\mathcal{L}$ that interlink with a nodal ring have a
nontrivial Berry bundle, which results in a nonzero topological
charge $\nu=1$. The nonzero topological charge ensures that
the nodal-ring cannot be gapped out by weak perturbations that preserve the symmetries.
Here, we should notice that both type-I and type-II nodal-rings share
the same protection mechanisms, although
there may not be a global gap along the loop $\mathcal{L}$ for the type-II case.
In Sec.\ref{field}, the topological response for the system is discussed, and we
reveal that the effective action near the the nodal-line is Chern-Simons type
for both type-I and type-II NLSM phases.

\section{\label{section3}Topological responses and the anomalous Hall conductivity}

\begin{figure}
  \centering
  \includegraphics[width=2.9in]{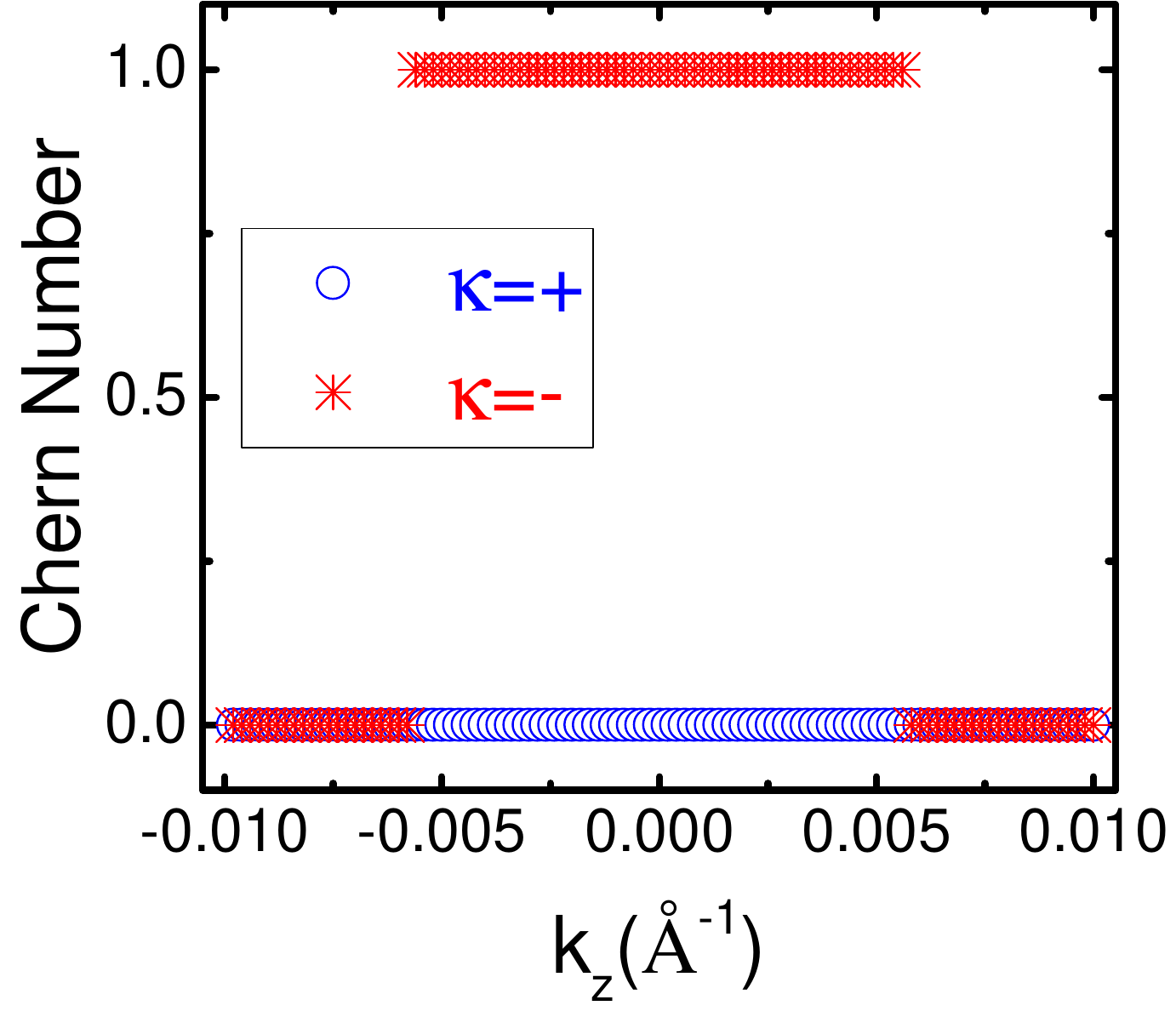}
  \caption{Plot of the Chern numbers as functions of $k_z$,
  here we take $g_{\mathrm{z}}=0.116$eV
  and $A_2=1.853$eV$\cdot${\AA}, other parameters are the same as Fig.\ref{Band}.}\label{Chern}
\end{figure}

The topological nature for topological semimetals is related to
unusual electromagnetic response characteristics,
and can be probed via topological transport phenomena, for example, the Hall conductivities.
For the system we considered above,
TRS is broken by the exchange field,
so the anomalous Hall effect is induced.
Here, we use the standard field theoretical approach to calculate
the response function~\cite{Fradkin}, from which we can obtain the
Hall conductivities and the effective action in a systematic way.

We start from introducing the (3+1)-dimensional U(1) gauge field
$A_\mu=(A_0,\mathbf{A})$ to couple the fermion, and
the action can be written as
\begin{equation}
\begin{split}
&S[\bar{\psi},\psi,A_\mu]\\&=\int\mathrm{d}^3\mathbf{x}
\mathrm{d}\tau\,\bar{\psi}_{\mathbf{x}}(\tau)
[-\partial_\tau-\mathrm{i}eA_0+\mathcal{H}(-\mathrm{i}\partial_i-eA_i)]\psi_{\mathbf{x}}(\tau)\ ,
\end{split}
\end{equation}
where the Hamiltonian is minimally coupled to the vector potential $\mathbf{A}$.
Performing the Fourier transformation to the four-momentum space and expanding the Hamiltonian
to linear order of $\mathbf{A}$, we get
\[
H\simeq\sum_{\mathbf{k},n}\psi^\dag_{\mathbf{k}n}\mathcal{H}(\mathbf{k})\psi_{\mathbf{k}n}
+\sum_{\mathbf{q}}\mathbf{J}_{\mathbf{q}}\cdot\mathbf{A}(-\mathbf{q})\ ,
\]
where $\mathbf{J}_{\mathbf{q}}=\sum_{\mathbf{k},n}\psi^\dag_{\mathbf{k}+\mathbf{q}/2,n}
\partial_\mathbf{k}\mathcal{H}\psi_{\mathbf{k}-\mathbf{q}/2,n}$ is the current operator.
The action becomes
\begin{equation}
\begin{split}
S[\bar{\psi},\psi,A_\mu]&=\sum_{\mathbf{k}}\bar{\psi}_{\mathbf{k},n}
[\mathrm{i}\omega_n-\mathrm{i}eA_0-\mathcal{H}(\mathbf{k})]\psi_{\mathbf{k},n}
\\&\quad-\sum_{\mathbf{k},\mathbf{q},n}
\bar{\psi}_{\mathbf{k}+\mathbf{q}/2,n}
\partial_\mathbf{k}\mathcal{H}\psi_{\mathbf{k}-\mathbf{q}/2,n}
\cdot\mathbf{A}(-\mathbf{q})\ .
\end{split}
\end{equation}
The partition function for the system reads
$\mathcal{Z}=\int D[A_\mu]\mathcal{Z}[A_\mu]$,
where $\mathcal{Z}[A_\mu]\equiv\mathrm{e}^{-S_{\mathrm{eff}}[A_\mu]}$
defines effective action for the U(1) gauge field $A_\mu$,
which can be obtained by
integrating out the fermion fields,
\begin{equation}
\mathrm{e}^{-S_{\mathrm{eff}}[A_\mu]}
=\int D[\bar{\psi},\psi]\,\mathrm{e}^{-S[\bar{\psi},\psi,A_\mu]}\ .
\end{equation}
The effective action $S_{\mathrm{eff}}[A_\mu]$ determines the response of the system
to external electromagnetic field. More explicitly, the expectation value of the
current is given by $\langle J^\mu\rangle=\delta S_{\mathrm{eff}}/\delta A_{\mu}$.
Expanding the effective action in the powers of $A_\mu$ to the second order,
we get
\begin{equation}
S^{(2)}_{\mathrm{eff}}[A_\mu]=\frac{1}{2}\int\frac{\mathrm{d}^3q}{(2\pi)^3}\Pi^{\mu\nu}(q)A_\mu(q)A_\nu(-q)\ ,
\end{equation}
where $\Pi_{\mu\nu}(q)$ is the current--current correlation function
\begin{equation}
\begin{split}
\Pi_{\mu\nu}(q)&=\langle J^{\mu}_{-\mathbf{q}}J^{\nu}_{\mathbf{q}}\rangle
\\&=\int\frac{\mathrm{d}^3k}{(2\pi)^3}\mathrm{tr}\big(\partial_{k_\mu}\mathcal{H}
G_{n+m,\mathbf{k}+\mathbf{q}}\partial_{k_\nu}\mathcal{H}G_{n,\mathbf{k}}\big)\ ,
\end{split}
\end{equation}
where
$k=(\mathrm{i}\omega_n,\mathbf{k})$ and $q=(\mathrm{i}\Omega_m,\mathbf{q})$
are the four-component momentum with $\omega_n$ and $\Omega_m$
denoting fermionic and bosonic Matsubara frequency, respectively.
$G_k$ is the Matsubara Green's function, whose form is given by
\begin{equation}
G_{n,\mathbf{k}}\equiv-\langle\psi_{\mathbf{k},n}\bar{\psi}_{\mathbf{k},n}\rangle
=\frac{1}{\mathrm{i}\omega_n-\mathcal{H}(\mathbf{k})}\ .
\end{equation}
The interesting component of the Hall conductivity is $\sigma_{xy}$,
which can be obtained through the relation
\begin{equation}
\sigma_{xy}=\lim_{q\rightarrow0}\frac{\Pi_{xy}(\mathrm{i}\Omega_m\rightarrow\Omega+\mathrm{i}0^+)}{-\mathrm{i}\Omega}\ .
\end{equation}
It is straightforward to verify that the Hall conductivity does not
depend on the kinetic energy $\epsilon_{\mathbf{k}}$ (except through the Fermi functions).
The calculation details can be found in the Appendix \ref{appA}.
We hence obtain, for the Hall conductivity, 
\begin{equation}
\sigma_{xy}=\frac{e^2}{\hbar}\int_{\mathrm{BZ}}\frac{\mathrm{d}^3k}{(2\pi)^3}
\sum_{\mu,\kappa} n_{\mathrm{F}}[E^\mu_\kappa(\mathbf{k})]F^{\mu}_{\kappa,xy}(\mathbf{k})\ ,
\label{HallConduc}
\end{equation}
where $F^{\mu}_{\kappa,xy}(\mathbf{k})=\mathrm{i}\hat{\mathbf{e}}_z\cdot
[\nabla_\parallel\times\langle\psi^{\mu}_\kappa(\mathbf{k})
|\nabla_\parallel|\psi^{\mu}_\kappa(\mathbf{k})\rangle]$
is the Berry curvature, and $n_{\mathrm{F}}[E^\mu_\kappa(\mathbf{k})]$
represents the Fermi distribution function.

In Figs.\ref{Berry}(b)-(d),
we illustrate the configurations of the Berry curvature
for subbands with $\kappa=-$ in momentum space.
Fig.\ref{Berry}(b) shows the case that $g_{\mathrm{z}}<|M_0|$,
the system is in the trivial insulator phase, one can find
there exists a net Berry curvature flow along the $z$-axis, such flow
will induce a negative magnetoresistance, even in the absence of chiral anomaly, or Weyl nodes~\cite{Lu}.
In Fig.\ref{Berry}(c), we switch on a small $\mathcal{PT}$-symmetry breaking term $\Delta$,
which induces circular Berry curvature flows in $xy$-plane.
This gives rise to a transverse Hall current, given by~\cite{Xiao1,Xiao2}
\begin{equation}
\mathbf{j}_{\mathrm{t}}=\frac{e^2}{\hbar}\int\frac{\mathrm{d}^3k}{(2\pi)^3}
n_{\mathrm{F}}[E(\mathbf{k})]\mathbf{E}\times\mathbf{F}(\mathbf{k})\ ,
\end{equation}
where $\mathbf{E}$ represents the electric field.
When $g_{\mathrm{z}}>|M_0|$,
we can find two monopoles in Fig.\ref{Berry}d,
corresponding to the Weyl semimetal phase.

\begin{figure}
  \centering
  \includegraphics[width=2.8in]{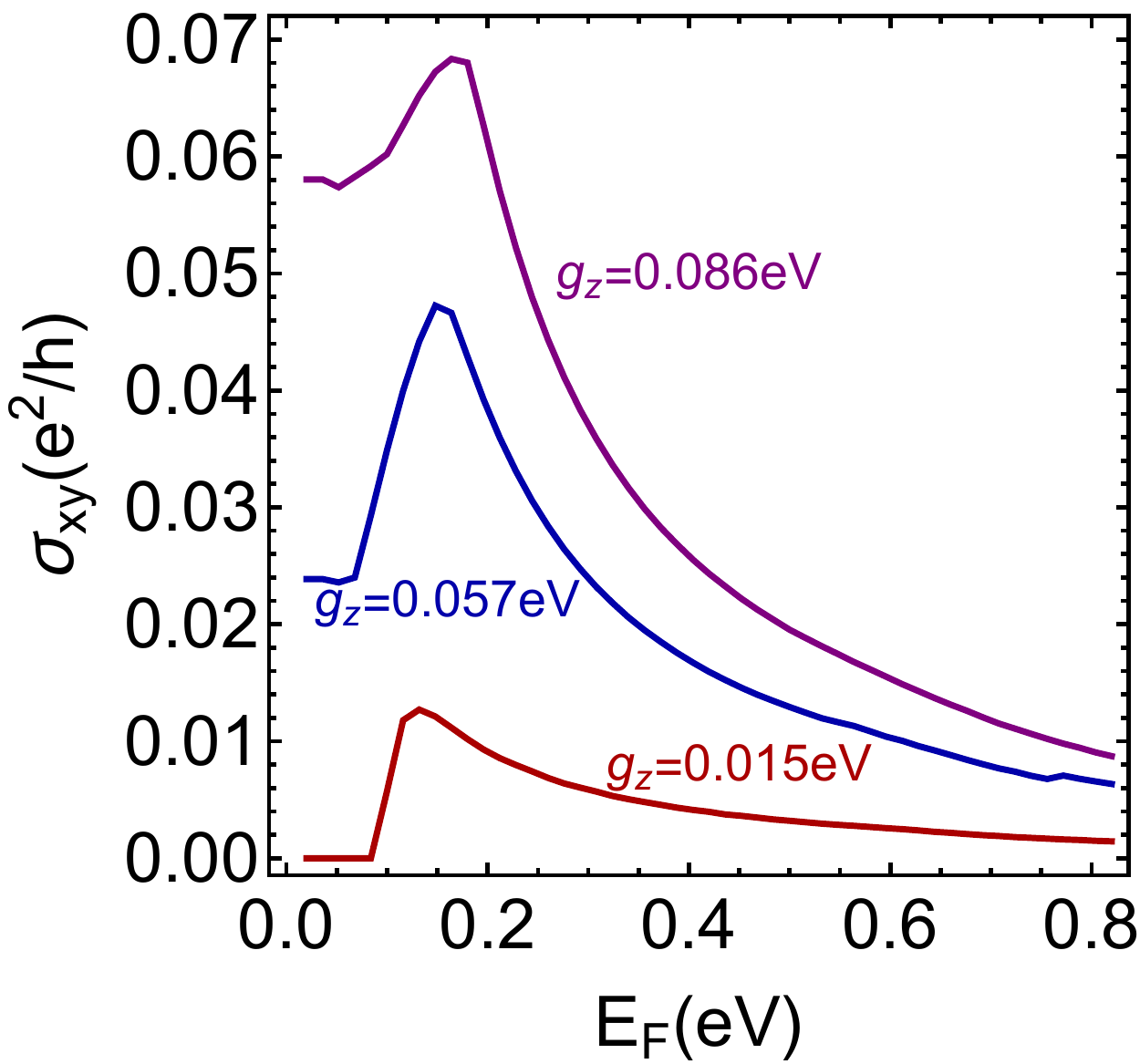}
  \caption{Plot of the Hall conductivity $\sigma_{xy}$ as a function of Fermi energy $E_{\mathrm{F}}$
  with different values of Zeeman coupling strength $g_\mathrm{z}$.
  Here, $A_2=1.853$eV$\cdot${\AA}, other parameters are the same as Fig.\ref{Band}.}
  \label{conductivity}
\end{figure}

It is convenient to put the integration in Eq.(\ref{HallConduc})
into the cylinder coordinate system, where the gradient operator can be written as
$\nabla_\parallel=\hat{\mathbf{e}}_{k_\parallel}\partial_{k_\parallel}+\hat{\mathbf{e}}
_{\varphi_{\mathbf{k}_\parallel}}(1/k_\parallel)\partial_{\varphi_{\mathbf{k}_\parallel}}$.
Noticing that $\partial_{k_\parallel}\hat{\mathbf{e}}_{\varphi_{\mathbf{k}_\parallel}}=0$ and
$\partial_{\varphi_{\mathbf{k}_\parallel}}\hat{\mathbf{e}}_{\varphi_{\mathbf{k}_\parallel}}=
-\hat{\mathbf{e}}_{k_\parallel}$, a straightforward algebraic manipulation yields
\begin{equation}
\sigma_{xy}=\frac{e^2}{h}\int_{\mathrm{BZ}}\frac{\mathrm{d}k_z}{2\pi}\int k_\parallel \mathrm{d}k_\parallel
\sum_{\mu,\kappa}n_{\mathrm{F}}[E^\mu_\kappa]F^{\mu}_{\kappa}(k_\parallel,k_z)\ ,
\end{equation}
where $F^{\mu}_{\kappa}(k_\parallel,k_z)=\frac{1}{2k_\parallel}\frac{\partial}{\partial k_\parallel}
P^{\mu}_{\kappa}(k_\parallel,k_z)$ with
\begin{equation}
P^{\mu}_\kappa(k_\parallel,k_z)=2\mathrm{i}\langle\psi^{\mu}_\kappa(\mathbf{k})|\partial_{\varphi_{\mathbf{k}_\parallel}}
|\psi^{\mu}_\kappa(\mathbf{k})\rangle\ .
\end{equation}
At zero temperature, the Fermi distribution function simply takes the form $n_{\mathrm{F}}[E^\mu_\kappa(k_\parallel,k_z)]
=\Theta[E_{\mathrm{F}}-E^\mu_\kappa(k_\parallel,k_z)]$, where $\Theta$ is the Heaviside
step function. As a typical situation, we first set the Fermi level at the energy of the two Weyl nodes,
for which only valence bands contribute to the Hall conductivity. When $k_z$ is treated as a parameter,
the system we considered is equivalent to a 2D Chern insulator.
The Hall conductivity
for the fully occupied valence bands can be derived to be
\begin{equation}
\sigma^{\mathrm{v}}_{xy}=\frac{e^2}{h}\int_{\mathrm{BZ}}\mathrm{d}k_z\sum_\kappa C_\kappa(k_z)
=2k^{\mathrm{w}}_z\frac{e^2}{h}\ ,
\label{cond1}
\end{equation}
where
\begin{equation}
C_\kappa(k_z)=\frac{1}{2}\big(\mathrm{sgn}[\Delta_\kappa(k_z)]-\mathrm{sgn}[\kappa]\big)
\end{equation}
is the first Chern number and $\Delta_\kappa(k_z)=\Delta_\kappa(k_\parallel=0,k_z)$.
We can find that the Hall conductivity for fully occupied valence bands is proportional
to the distance between the two Weyl nodes $(0,0,\pm k^{\mathrm{w}}_z)$, as expected, and
only the subbands with $\kappa=-$ carry nonzero Chern numbers
in the region of $k_z$ between Weyl nodes, as Fig. \ref{Chern} shows, 
whereas the subbands with $\kappa=+$
do not contribute to the Hall conductivity as long as the Fermi level stays away from them,
this is known as the Weyl metal phase~\cite{burkov}.
As long as the Fermi energy starts merging into conduction bands,
the anomalous Hall conductivity is tuned
due to the Fermi surface contribution,
whose form is derived to be
\begin{equation}
\sigma^{\mathrm{c}}_{xy}(E_{\mathrm{F}})=\frac{e^2}{2h}\int_{\mathrm{BZ}}\mathrm{d}k_z
\sum_\kappa\bigg(\Gamma_\kappa(k_z)-\mathrm{sgn}\big[\Delta_\kappa(k_z)\big]\bigg)\ ,
\end{equation}
where
\[
\Gamma_\kappa(k_z)\equiv\Gamma_\kappa(k=k_{\mathrm{F}},k_z)=
\cos2\alpha_{\mathbf{k}\kappa}\big|_{k_\parallel=\sqrt{k^2_{\mathrm{F}}-k^2_z}}
\]
is defined at the Fermi energy $E_{\mathrm{F}}$.
The total anomalous Hall conductivity can be given as
\begin{equation}
\begin{split}
\sigma_{xy}&=\sigma^{\mathrm{v}}_{xy}+\sigma^{\mathrm{c}}_{xy}(E_{\mathrm{F}})
\\&=\frac{e^2}{2h}\int_{\mathrm{BZ}}\frac{\mathrm{d}k_z}{2\pi}\sum_\kappa
\Big[\Gamma_\kappa(k_z)-\mathrm{sgn}(\kappa)\Big]\ .
\end{split}
\label{cond2}
\end{equation}
In Fig.\ref{conductivity}, we numerically illustrate the relations between $\sigma_{xy}$
and the Fermi energy $E_{\mathrm{F}}$ under different Zeeman coupling strengths.
We can find that when $g_{\mathrm{z}}<|M_0|$,  the system is in the trivial insulator phase,
thus $\sigma^{\mathrm{v}}_{xy}=0$ when the Fermi energy stays in the band gap.
When $g_{\mathrm{z}}>|M_0|$, the system is in the Weyl semimetal phase,
then $\sigma^{\mathrm{v}}_{xy}$ is proportional to
the distance between the two Weyl nodes, as Eq.(\ref{cond1}) shows.
Further, with the increasing of the Fermi energy, the subband with $\kappa=+1$
contributes a negative term in the total Berry curvature, leading the decrease of $\sigma_{xy}$
at large Fermi energies.

\section{Effective action close to the nodal-line}\label{field}
In this section, we derive the effective action for the electron states
which are very close to the nodal-lines, i.e. only their low energy physics is considered.
As discussed in Sec.\ref{section2},
the low energy effective Hamiltonian for NLSM phase is defined in a
(2+1)-dimensional subspace parameterized by $\phi$,
which characterizes a pair of massless Dirac fermion.
The parity anomaly Chern-Simons action is obtained after introducing the
mass term.

Introducing the (2+1)-dimensional gauge
potential $A_\mu=(A_0,A_\parallel,A_z)$, we can
write the action functional for a given nodal point $\chi$ as
\begin{equation}
\begin{split}
&S^\chi[\bar{\psi},\psi,A_\mu]\\&=\int\mathrm{d}^2\mathbf{x}
\mathrm{d}\tau\,\bar{\psi}^\chi_{\mathbf{x}}(\tau)
[-\partial_\tau-\mathrm{i}eA_0+\mathcal{H}^\chi_{\mathrm{NL}}
(-\mathrm{i}\partial_i-eA_i)]\psi^\chi_{\mathbf{x}}(\tau)\ ,
\end{split}
\end{equation}
here, $\mathbf{x}=(r_\parallel,z)$.
To obtain the topological response theory,
we should introduce a pseudo $\mathcal{PT}$-symmetry
breaking term $\Delta\hat{\tau}_y\hat{\sigma}_z$ in $\mathcal{H}^\chi_{\mathrm{NL}}$
for regularization
due to the ultra-violet divergences of the massless Dirac fermions~\cite{reg1}.
This term usually can be induced by the spin-orbit coupling
or the trigonal momentum correction term for the effective Hamiltonian~\cite{model}.
Repeating the calculations we have done in Sec.\ref{section3},
we can obtain an effective action $S^\chi_{\mathrm{eff}}[A_\mu]$.
Expanding $S^\chi_{\mathrm{eff}}[A_\mu]$ in powers of $A_\mu$ to the second order, we get
\begin{equation}
S^{\chi,(2)}_{\mathrm{eff}}[A_\mu]=\frac{1}{2}\int\frac{\mathrm{d}^3q}{(2\pi)^3}\Pi^\chi_{\mu\nu}(q)A_\mu(q)A_\nu(-q)\ ,
\end{equation}
where
\begin{equation}
\Pi^\chi_{\mu\nu}(q)=\int\frac{\mathrm{d}^3p}{(2\pi)^3}\mathrm{tr}\big[J^\chi_\mu G^\chi(p+q)J^\chi_\nu G^\chi(p)\big]
\end{equation}
is the correlation function, and $J^\chi_\mu=\partial_{k_\mu}\mathcal{H}^\chi_{\mathrm{NL}}$.
Expanding the correlation function to linear order of $q$
(the long wavelength limit $q\rightarrow0$) at the zero temperature,
we finally arrive at the Chern-Simons action
\begin{equation}
S^{\chi,(2)}_{\mathrm{eff}}[A_\mu]=\frac{\mathrm{sgn}(\mu\chi\Delta)}{8\pi}\int\mathrm{d}^2x\mathrm{d}t\,
\epsilon^{\mu\nu\tau}A_\mu\partial_\nu A_\tau\ ,
\label{cs}
\end{equation}
where $\mu=+1(-1)$ for conduction (valence) band.
Equation (\ref{cs}) confirms that
the Chern-Simons term is independent of parameters $g_{\mathrm{z}}$ and $A_2$,
hence it can describe parity anomalies for both type-I and type-II NLSMs.
The coefficients $\mathrm{sgn}(\mu\chi\Delta)$ for different values of $\mu$ and $\chi$
are in fact the $\mathbb{Z}_2$ topological charges,
as they emerge in pairs and
the summation over all of the topological charges is zero.
This reflects the $\mathbb{Z}_2$ nature of the parity anomaly
in the $\tau$-subspaces.
This is an exotic feature of NLSMs,
because the parity anomaly usually only occurs in (2+1)-dimensional systems,
and now appears in the (3+1)-dimensional NLSMs.
However, the conductivity for the parity anomaly NLSM also becomes zero,
which makes the parity anomaly hard to be measured.
In a recent work~\cite{Matsushita}, authors proposed that the parity anomaly in NLSMs
can be measured by the topological piezoelectric effect, which can be
realized via periodic lattice deformations.

\section{Conclusion}
We theoretically studied the topological phases in a magnetic three-dimensional
topological insulator in the low-energy description. We found that both the
Weyl nodes and nodal-line can emerge and coexist with
the increasing of the Zeeman exchange strength,
in which both type-I and type-II nodal-line can be obtained by tuning some parameters.
In addition, we analyzed the topological responses near the nodal-line
and obtained the effective Chern-Simons action.
We also computed the anomalous Hall conductivity for the full Hamiltonian.

\begin{acknowledgments}
M. N. C would like to thank Wei Su and Wei Chen for helpful discussions.
This work was supported by the National Natural Science Foundation of China
under grant numbers 11804070 (MNC) and 61805062 (YZ).
\end{acknowledgments}

\appendix
\section{Derivation od the Chern-Simons effective action}\label{appA}
In this section we give the derivation of the Chern-Simons effective action.
We start from the low-energy effective Hamiltonian given by Eq.(\ref{nlsm1}) in the main text.
Introducing a pseudo $\mathcal{PT}$-symmetry breaking term $\Delta\hat{\tau}_y\hat{\sigma}_z$,
the effective Hamiltonian becomes
\begin{equation}
\begin{split}
\mathcal{H}^\chi_{\mathrm{NL}}(\mathbf{p})&=A_2p_\parallel\hat{\tau}_0(\cos\phi\hat{\sigma}_x+\sin\phi\hat{\sigma}_y)
+g_\mathrm{z}\hat{\sigma}_z\\&\quad+(-\chi A_3p_\parallel\hat{\tau}_z+A_1p_z\hat{\tau}_x+\Delta\hat{\tau}_y)\hat{\sigma}_z\ .
\end{split}
\label{nlsm2}
\end{equation}
We can diagonalize Eq.(\ref{nlsm2}) to obtain the eigenvalues:
\begin{equation}
E_{\kappa\pm}(p)=\pm\sqrt{A^2_2p^2_\parallel+(g_z+d_\kappa)^2}\ ,
\end{equation}
and their corresponding eigenstates
$|\Psi^{\chi}_{\kappa\mu}(\mathbf{p})\rangle
=|\phi^\chi_\kappa(\mathbf{p})\rangle\otimes|\psi_{\kappa\mu}(\mathbf{p})\rangle$,
where $d_\kappa=\kappa\sqrt{A^2_3p^2_\parallel+A^2_1p^2_z+\Delta^2}$ with $\kappa=\pm1$,
\begin{equation}
\begin{split}
|\phi^\chi_+(\mathbf{p})\rangle&=\begin{pmatrix}\mathrm{e}^{-\mathrm{i}\gamma_{\mathbf{p}}}\cos\varphi^\chi_{\mathbf{p}}\\
-\sin\varphi^\chi_{\mathbf{p}}\end{pmatrix}\ , \\
|\phi^\chi_-(\mathbf{p})\rangle&=\begin{pmatrix}\mathrm{e}^{-\mathrm{i}\gamma_{\mathbf{p}}}\sin\varphi^\chi_{\mathbf{p}}\\
\cos\varphi^\chi_{\mathbf{p}}\end{pmatrix}\ ,
\end{split}
\end{equation}
and
\begin{equation}
\begin{split}
|\psi^{\kappa}_+(\mathbf{p})\rangle&=\begin{pmatrix}\mathrm{e}^{-\mathrm{i}\phi}\sin\alpha^\kappa_{\mathbf{p}}
\\-\cos\alpha^\kappa_{\mathbf{p}}\end{pmatrix}\ , \\
|\psi^{\kappa}_-(\mathbf{p})\rangle&=\begin{pmatrix}\mathrm{e}^{-\mathrm{i}\phi}\cos\alpha^\kappa_{\mathbf{p}}\\ \sin\alpha^\kappa_{\mathbf{p}}\end{pmatrix}\ .
\end{split}
\end{equation}
In above expressions, we have set parameters to simplify the results, namely:
\[
\gamma_{\mathbf{p}}=\mathrm{Arg}\frac{A_1p_z+\mathrm{i}\Delta}{\sqrt{A^2_1p^2_z+\Delta^2}}\ , 
\]
\[
2\varphi^\chi_{\mathbf{p}}=\mathrm{arccot}\frac{\chi A_3p_\parallel}{\sqrt{A^2_1p^2_z+\Delta^2}}\ ,
\]
\[
2\alpha^\kappa_{\mathbf{p}}=\arccos\frac{d_\kappa}{\sqrt{d^2_\kappa+A^2_2p^2_\parallel}}\ .
\]
The Green's function can now be written in the form
\begin{equation}
\hat{G}^\chi(\mathbf{p})=\sum_{\mu,\kappa}\frac{\hat{P}^{\chi}_{\kappa\mu}(\mathbf{p})}
{\mathrm{i}\omega_n-E_{\kappa\mu}(\mathbf{p})}
\end{equation}
where
\begin{equation}
\hat{P}^{\chi}_{\kappa\mu}(\mathbf{p})=|\Psi^{\chi}_{\kappa\mu}(\mathbf{p})
\rangle\langle\Psi^{\chi}_{\kappa\mu}(\mathbf{p})|
\end{equation}
defines the projection operators. With these results, now we can compute the
correlation function $\Pi^\chi_{\mu\nu}(q)$.
Here, we are only interested in the $0\alpha$-component, where $\alpha=\parallel,z$, then
\begin{widetext}
\[
\begin{split}
&\quad\mathrm{tr}\big[J^\chi_0 G^\chi(\mathbf{p}+\mathbf{q})J^\chi_\alpha G^\chi(\mathbf{p})\big]
\\&=\sum_{\mu,\nu,\kappa,\vartheta}\frac{1}{[\mathrm{i}(\omega_n+\Omega)-E_{\kappa\mu}(\mathbf{p}+\mathbf{q})]
[\mathrm{i}\omega_n-E_{\vartheta\nu}(\mathbf{p})]}
\mathrm{tr}\big[J^\chi_0\hat{P}^{\chi}_{\kappa\mu}(\mathbf{p} +\mathbf{q})J^\chi_\alpha\hat{P}^{\chi}_{\vartheta\nu}(\mathbf{p})\big]
\\&=\sum_{\mu,\nu,\kappa,\vartheta}\bigg[\frac{1}{\mathrm{i}\omega_n-E_{\vartheta\nu}(\mathbf{p})}
-\frac{1}{\mathrm{i}(\omega_n+\Omega)-E_{\kappa\mu}(\mathbf{p}+\mathbf{q})}\bigg]\frac{1}{\mathrm{i}\Omega
-[E_{\kappa\mu}(\mathbf{p}+\mathbf{q})-E_{\vartheta\nu}(\mathbf{p})]}
\mathrm{tr}\big[J^\chi_0\hat{P}^{\chi}_{\kappa\mu}(\mathbf{p}+\mathbf{q}) J^\chi_\alpha\hat{P}^{\chi}_{\vartheta\nu}(\mathbf{p})\big]
\end{split}
\]
After Matsubara frequency summation, we obtain
\begin{equation}
\begin{split}
\mathrm{tr}\big[J^\chi_0 G^\chi(\mathbf{p}+\mathbf{q})J^\chi_\alpha G^\chi(\mathbf{p})\big]&=
\sum_{\mu,\nu,\kappa,\vartheta}\frac{n_{\mathrm{F}}[E_{\kappa\mu}(\mathbf{p}+\mathbf{q})-\mathrm{i}\Omega]-n_{\mathrm{F}}[E_{\vartheta\nu}(\mathbf{p})]}
{\mathrm{i}\Omega-[E_{\kappa\mu}(\mathbf{p}+\mathbf{q})-E_{\vartheta\nu}(\mathbf{p})]}
\mathrm{tr}\big[J^\chi_0\hat{P}^{\chi}_{\kappa\mu}(\mathbf{p}+\mathbf{q}) J^\chi_\alpha\hat{P}^{\chi}_{\vartheta\nu}(\mathbf{p})\big]\ .
\end{split}
\end{equation}
Expanding to linear order of $q\equiv(\mathrm{i}\Omega,\mathbf{q})$ and take the DC limit $\Omega\rightarrow0$, we obtain
\[
\begin{split}
\mathrm{tr}\big[J^\chi_0 G^\chi(\mathbf{p}+\mathbf{q})J^\chi_\alpha G^\chi(\mathbf{p})\big]&\simeq
\sum_{\mu,\nu,\kappa,\vartheta}\frac{n_{\mathrm{F}}[E_{\kappa\mu}(\mathbf{p})]-n_{\mathrm{F}}[E_{\vartheta\nu}(\mathbf{p})]}
{-[E_{\kappa\mu}(\mathbf{p})-E_{\vartheta\nu}(\mathbf{p})]}\\&\quad\times
\langle\Psi^{\chi}_{\vartheta\nu}(\mathbf{p})|\Psi^{\chi}_{\kappa\mu}(\mathbf{p}+\mathbf{q})\rangle
\langle\Psi^{\chi}_{\kappa\mu}(\mathbf{p}+\mathbf{q})|J^\chi_\alpha|\Psi^{\chi}_{\vartheta\nu}(\mathbf{p})\rangle\ .
\end{split}
\]
Next, expanding the state vector to linear order of $\mathbf{q}$, we get
\[
\langle\Psi^{\chi}_{\vartheta\nu}(\mathbf{p})|\Psi^{\chi}_{\kappa\mu}(\mathbf{p}+\mathbf{q})\rangle\simeq
\delta_{\vartheta\kappa}\delta_{\mu\nu}+q_\beta
\langle\Psi^{\chi}_{\vartheta\nu}(\mathbf{p})|\partial_{p_\beta}|\Psi^{\chi}_{\kappa\mu}(\mathbf{p})\rangle\ .
\]
Using the relation that
\[
\langle\Psi^{\chi}_{\vartheta\nu}(\mathbf{p})|\partial_{p_\beta}|\Psi^{\chi}_{\kappa\mu}(\mathbf{p})\rangle
=\frac{\langle\Psi^{\chi}_{\vartheta\nu}(\mathbf{p})|
\frac{\partial \mathcal{H}^{\chi}_{\mathrm{NL}}}{\partial p_\beta}|\Psi^{\chi}_{\kappa\mu}(\mathbf{p})\rangle}
{-E_{\vartheta\nu}(\mathbf{p})+E_{\kappa\mu}(\mathbf{p})}\ ,
\]
taking the case $\mu=\nu$ because the nodal-line can only exist between two conduction bands and two valence bands,
and noticing that $\vartheta\neq\kappa$, so we obtain
\[
\begin{split}
&\quad\mathrm{tr}\big[J^\chi_0 G^\chi(\mathbf{p}+\mathbf{q})J^\chi_\alpha G^\chi(\mathbf{p})\big]\\&\simeq
\sum_\mu n_{\mathrm{F}}[E_{\kappa\mu}(\mathbf{p})]\frac{\langle\Psi^{\chi}_{\vartheta\mu}(\mathbf{p})|
\frac{\partial \mathcal{H}^{\chi}_{\mathrm{NL}}}{\partial p_\beta}|\Psi^{\chi}_{\kappa\mu}(\mathbf{p})\rangle
\langle\Psi^{\chi}_{\kappa\mu}(\mathbf{p})|\frac{\partial
\mathcal{H}^{\chi}_{\mathrm{NL}}}{\partial p_\alpha}|\Psi^{\chi}_{\vartheta\mu}(\mathbf{p})\rangle-(\vartheta\leftrightarrow\kappa)}
{[E_{\kappa\mu}(\mathbf{p})-E_{\vartheta\mu}(\mathbf{p})]^2}q_\beta\ .
\end{split}
\]
\end{widetext}
Therefore, we can find that the effective action has the form
$S^{\chi,(2)}_{\mathrm{eff}}[A_\mu]=(\mathcal{C}^\chi/2)\epsilon^{0\beta\alpha}A_0q_\beta A_\alpha$,
which is exactly the Chern-Simons type action. Next, we shall calculate the coefficient $\mathcal{C}^\chi$.
At zero temperature, $n_{\mathrm{F}}[E_{\kappa\mu}(\mathbf{p})]=1$ for all occupied states.
Therefore, the coefficient $\mathcal{C}^\chi$ becomes the Berry phase, which is given by
\begin{equation}
\mathcal{C}^\chi_{\alpha\beta}=\sum_{\mu\in\mathrm{occ.}}
\int\frac{\mathrm{d}^3p}{(2\pi)^3}F^\chi_{\alpha\beta}(\mathbf{p})\ ,
\end{equation}
where
\[
F^\chi_{\alpha\beta}(\mathbf{p})=\big[\langle\partial_\alpha\Psi^{\chi}_{\kappa\mu}(\mathbf{p})|
\partial_\beta\Psi^{\chi}_{\kappa\mu}(\mathbf{p})\rangle-(\alpha\leftrightarrow\beta)\big]
\]
is the Berry curvature.
Using the chain rule, we have
\begin{equation}
\langle\Psi^{\chi}_{\kappa\mu}|
\partial_\alpha\Psi^{\chi}_{\kappa\mu}\rangle=\sum_{\ell}A_\ell\frac{\mathrm{d}\ell}{\mathrm{d}p_\alpha}\ ,
\end{equation}
where $\ell=\gamma_{\mathbf{p}},\varphi^\chi_{\mathbf{p}}$ and $\alpha^\kappa_{\mathbf{p}}$.
It is quite straightforward to obtain that $A_{\varphi^\chi_{\mathbf{p}}}=A_{\alpha^\kappa_{\mathbf{p}}}=0$
and $A_{\gamma_{\mathbf{p}}}=\cos^2\varphi^\chi_{\mathbf{p}}$, thus
\begin{equation}
\langle\Psi^{\chi}_{\kappa\mu}|\partial_\alpha\Psi^{\chi}_{\kappa\mu}\rangle
=A_{\gamma_{\mathbf{p}}}\frac{\mathrm{d}\gamma_{\mathbf{p}}}{\mathrm{d}p_\alpha}
=A_{\gamma_{\mathbf{p}}}\frac{1}{\cos\gamma_{\mathbf{p}}}\frac{\mathrm{d}(\sin\gamma_{\mathbf{p}})}
{\mathrm{d}p_\alpha}\ .
\end{equation}
After some algebra, we can obtain
\begin{equation}
F^\chi_{\parallel z}(\mathbf{p})=\frac{\mu\chi A_1A_3\Delta}{2(A^2_3p^2_\parallel+A^2_1p^2_z+\Delta^2)^{3/2}}\ .
\end{equation}
Integration over $\mathbf{p}$ yielding:
\begin{equation}
\mathcal{C}^\chi_{\alpha\beta}=\frac{1}{4\pi}\sum_{\mu\in\mathrm{occ.}}\mathrm{sgn(\mu\chi\Delta)}\ ,
\end{equation}
so Eq.(\ref{cs}) is obtained.

\end{document}